\documentclass[%
 reprint,
superscriptaddress,
 amsmath,amssymb,
 aps, 
 prl,
]{revtex4-2}

\usepackage{graphicx}
\usepackage{dcolumn}
\usepackage{bm}

\begin{document}


\title{Imprinting spatial helicity structure of vector vortex beam on spin texture in semiconductors}

\author{Jun Ishihara}
\email{j.ishihara@rs.tus.ac.jp}
\affiliation{Department of Applied Physics, Tokyo University of Science, Tokyo 125-8585, Japan}%
\author{Takachika Mori}%
\affiliation{Department of Applied Physics, Tokyo University of Science, Tokyo 125-8585, Japan}%
\author{Takuya Suzuki}%
\affiliation{Department of Applied Physics, Tokyo University of Science, Tokyo 125-8585, Japan}%
\author{Sota Sato}%
\affiliation{Graduate School of Electrical and Electronic Engineering, Chiba University, Chiba 263-8522, Japan}%
\author{Ken Morita}%
\affiliation{Graduate School of Electrical and Electronic Engineering, Chiba University, Chiba 263-8522, Japan}%
\author{Makoto Kohda}%
\affiliation{Department of Materials Science, Tohoku University, Sendai 980-8579, Japan}%
\author{Yuzo Ohno}%
\affiliation{Graduate School of Pure and Applied Sciences, University of Tsukuba, Tsukuba 305-8573, Japan}%
\author{Kensuke Miyajima}%
\affiliation{Department of Applied Physics, Tokyo University of Science, Tokyo 125-8585, Japan}%

%
%
%
%
%

\begin{abstract}
We present the transfer of the spatially variant polarization of topologically structured light to the spatial spin texture in a semiconductor quantum well.
The electron spin texture, which is a circular pattern with repeating spin-up and spin-down states whose repetition rate is determined by the topological charge, is directly excited by a vector vortex beam with a spatial helicity structure.
The generated spin texture efficiently evolves into a helical spin wave pattern owing to the spin--orbit effective magnetic fields in the persistent spin helix state by controlling the spatial wave number of the excited spin mode.
By tuning the repetition length and azimuthal angle, we simultaneously generate helical spin waves with opposite phases by a single beam.
\end{abstract}

\maketitle



Spatial structures have attracted considerable attention because their spatial modes offer a new information basis, as represented by space division multiplexing in optical communications.
Topologically structured light has unique properties, such as a doughnut-shaped intensity profile, orbital angular momentum (OAM) originating from a spatial phase structure, and spatially variant polarization.
Its applications include super-resolution microscopy using doughnut-shaped light patterns \cite{Hell1994,Willing2007}, fabrication of chirality-dependent helical nanoneedles using an optical vortex \cite{Toyoda2012,Toyoda2013}, and optically induced rotational motion of nanoparticles by OAM \cite{ONeil2002,Curtis2003}.
Structured light with spatially variant polarization, such as azimuthal and radial beams, is a vector beam with nonseparable spatial and polarization degrees of freedom.
This coupled spatial--polarization mode improves optical communication and optical metrology \cite{Ndagano2018,Zhao2015,Milione2015,Berg-Johansen2015,Hu2019}.

The spatial structures of spins in semiconductors have also drawn attention recently.
A persistent spin helix (PSH), a helical spin texture in space, emerges when Rashba-type \cite{Rashba1960} and Dresselhaus-type \cite{Dresselhaus1955} spin--orbit (SO) interactions are tuned to equal strengths in a (001)-grown quantum well (QW) with a zinc blende structure \cite{Schliemann2003,Bernevig2006,Koralek2009}.
In the PSH state, the spatial spin texture is protected from dephasing caused by spin-independent scattering because momentum-dependent SO effective magnetic fields are unidirectional and have an exact SU(2) symmetry.
PSH can be applied not only in semiconductor-based spintronic devices using long-lived spin states but also in devices using spatial spin modes.
Studies have reported direct observations of the PSH texture \cite{Walser2012,Ishihara2013,Ishihara2014,Altmann2014,Salis2014,Ishihara2022}, modulation of PSH frequencies by gate voltages \cite{Kohda2012,Ishihara2013}, and long-distance transport of the PSH by drift currents \cite{Kunihashi2016,Altmann2016,Passmann2018,Anghel2018,Anghel2020}.

\begin{figure}[b]
\includegraphics[scale=0.5]{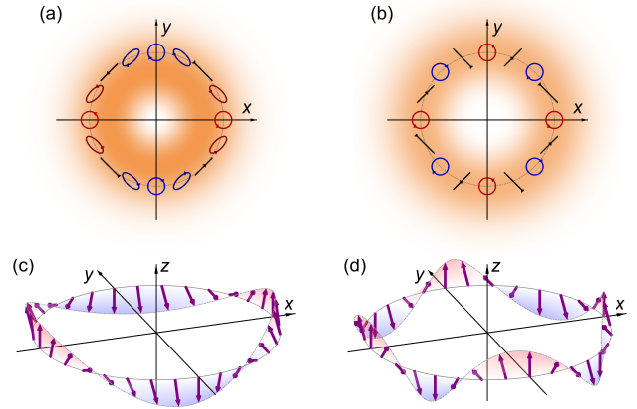}
\caption{\label{fig:1} Sketch of imprinting of polarization distribution by vector vortex beam on spin texture in semiconductor quantum well (QW).
Azimuthal angle dependence of polarization of vector vortex beam with (a) $l = 1$ and (b) $l = 2$.
(c), (d) Spatial spin structures photoexcited by vector vortex beam shown in (a) and (b) following optical selection rules.
The arrows represent the polarization directions in coherent transcription, and the colors represent the spin $z$ component.}
\end{figure}

Semiconductors have an excellent response to light, and spin-polarized electrons can be generated in semiconductors by polarized light following optical selection rules.
In addition to the optical generation and detection of spin populations, the coherent transfer of arbitrary light polarization to electron spins and the direct readout of coherent spin states have been demonstrated \cite{Kosaka2008,Kosaka2009}.
However, previous studies on optical spin injection used homogeneously polarized beams and did not focus on spatially structured light.
The use of vector vortex beams with spatial helicity structures is expected to generate spatial spin structures directly, thereby facilitating access to spatial spin modes and information conversion between structured light and spin textures.
Herein, we demonstrate the imprinting of spatial helicity structure of a vector vortex beam on the spin texture of electrons in a semiconductor QW.
A spatially periodic structure of the spin state is directly generated by the vector vortex beam.
In this structure, the spin-up and spin-down states are repeated in a circular pattern, and the structure period varies with the topological charge of the vector vortex beam.
The excitation of the periodic spin texture by the vector vortex beam corresponds to the excitation of a specific spatial spin wave mode.
This is in contrast to spin excitation by a conventional homogeneously polarized Gaussian beam, which excites various spin wave modes simultaneously.
The spatiotemporal dynamics of the spin texture reveals that the excited spin texture rapidly evolves into a PSH texture through the matching of the spatial spin mode excited by the vector vortex beam with the eigenmode of the helical spin wave of PSH induced by the SO effective magnetic field.
We also show that two PSH wave modes with opposite phases can be produced by a single vector vortex beam.

A vector vortex beam with an asymmetric helicity distribution is generated from a homogeneously polarized beam.
A linearly polarized beam is converted into an azimuthal beam by a vortex half-wave plate, whose fast axis rotates continuously in space while having a constant retardance of $\pi$.
Subsequently, the beam passes through a quarter-wave plate, and a spatial pattern of polarization is obtained.
The polarization of the vector vortex beam is expressed using the Jones vector as
\begin{eqnarray}
&&{\bf A}_\text{QWP}{\bf A}_\text{VHWP}{\bf J}_\text{V}\nonumber
\\&&=\frac{1}{\sqrt{2}}
\begin{pmatrix}
1 & i\\
i & 1
\end{pmatrix}
\begin{pmatrix}
\cos(l\phi) & -\sin(l\phi)\\
\sin(l\phi) & \cos(l\phi)
\end{pmatrix}
\begin{pmatrix}
0 \\
1
\end{pmatrix}\nonumber
\\&&=\frac{i}{\sqrt{2}}e^{il\phi}
\begin{pmatrix}
1 \\
-ie^{-i2l\phi}
\end{pmatrix}
\label{eq:1},
\end{eqnarray}
where ${\bf A}_\text{QWP}$ and ${\bf A}_\text{VHWP}$ are the Jones matrices of a quarter-wave plate and vortex half-wave plate, respectively, and ${\bf J}_\text{V}$ is the initial linear polarization.
The phase varies spatially with the azimuthal angle $\phi$, as represented by $e^{il\phi}$, thereby forming a vortex beam that carries an OAM determined by the topological charge $l$.
The spatial distribution of polarization forms a periodic structure of helicity, where left- and right-circularly polarized lights are repeated in a circular pattern, as shown in Figs.~\ref{fig:1}(a) and \ref{fig:1}(b).
The vector vortex beam expressed by Eq.~(\ref{eq:1}), as well as an optical vortex and the azimuthal beam, is also a polarization state that can be described by a higher-order Poincar\'{e} sphere \cite{Ren2015,Morita2022}, which is an extension of the Poincar\'{e} sphere representing the circular polarization state.
When the light polarization of the vector vortex beam is coherently transferred to the electron spins, the spin states are expected to have an azimuthal angle-dependent spatial structure, as shown in Figs.~\ref{fig:1}(c) and \ref{fig:1}(d). The arrows and the blue and red shading in Figs.~\ref{fig:1}(c) and \ref{fig:1}(d) represent the direction of spin polarization and the out-of-plane spin component, respectively.

The sample structure in this study is a one-sided modulation-doped 20 nm GaAs/AlGaAs QW grown on a semi-insulating (001) GaAs substrate by molecular beam epitaxy.
A 20 nm thick high Si-doped AlGaAs layer is placed on top of the QW with an undoped 35 nm thick AlGaAs spacer layer, providing high-mobility two-dimensional electron gas (2DEG).
The carrier density $n$ is $3.5\times10^{15}\ \text{m}^{-2}$; it is determined by the Stokes shift energy, which is obtained from photoluminescence and photoluminescence excitation spectra.
This quantum structure induces a strong Rashba SO interaction via the asymmetric band profile of the QW, resulting in SO fields close to the PSH.
The time evolution of the spin distribution is measured by time-resolved, spatially resolved magneto-optical Kerr rotation (KR).
The pulse train, which is generated by a mode-locked Ti:sapphire laser with a repetition rate of 82 MHz and a pulse duration of $\sim$1.5 ps, is split into pump and probe beams.
These beams are focused on the sample surface by an objective lens.
The relative delay time and the relative position between the pump and probe beams are automatically controlled by a mechanical delay line and a scanning mirror, respectively.
For the pump beam, a linearly polarized Gaussian beam is converted into doughnut-shaped, azimuthally polarized beam by a liquid crystal vortex half-wave plate.
Subsequently, the helicity distribution of circularly polarized beam, expressed by Eq.~(\ref{eq:1}), is generated by a photoelastic modulator.
The helicity is repeated with left- and right-circularly polarized beams at 50.147 kHz.
The polarized electron spins are optically excited by the pump beam following the optical selection rule, and the out-of-plane spin component is detected as a KR signal of the linearly polarized probe beam.
Optical measurements are performed at 7.5 K using a cryostat with optical access.

\begin{figure}[b]
\includegraphics[scale=1]{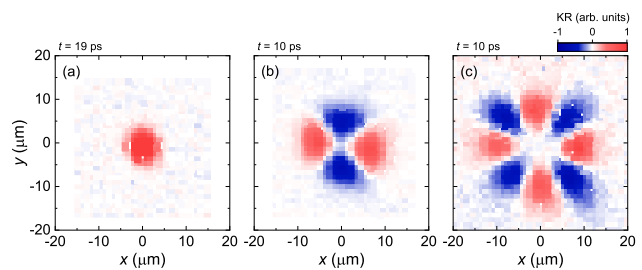}
\caption{\label{fig:2} KR maps corresponding to spatial spin distribution $s_z(x,y)$.
(a) $s_z(x,y)$ at $t=19$ ps excited by homogeneously polarized beam.
(b) $s_z(x,y)$ at $t=10$ ps excited by vector vortex beam with $l = 1$.
(c) $s_z(x,y)$ at $t=10$ ps excited by vector vortex beam with $l = 2$.}
\end{figure}

\begin{figure*}[t]
\includegraphics[scale=1]{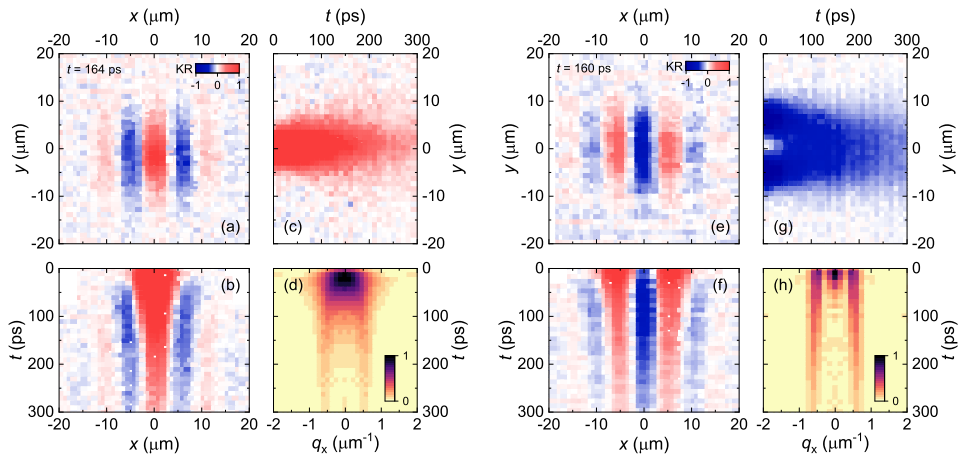}
\caption{\label{fig:3} (a) Spatial spin distribution $s_z(x,y)$ at $t = 164$ ps.
(b) Spatiotemporal map of $x$ direction $s_z(x,0,t)$.
(c) Spatiotemporal map of $y$ direction $s_z(0,y,t)$.
(d) Time evolution of spin wave mode distribution $A(q_x,t)$, obtained by frequency analysis of (b) under homogeneously polarized beam spin excitation.
(e) $s_z(x,y)$ at $t = 160$ ps, (f) $s_z(x,0,t)$, (g) $s_z(0,y,t)$, and (h) $A(q_x,t)$ obtained from (f) under vector vortex beam spin excitation.}
\end{figure*}

Figures~\ref{fig:2}(a)-\ref{fig:2}(c) show KR maps corresponding to the spin distribution $s_z(x,y)$ immediately after spin excitation under the different pump beams.
Electron spins excited by a homogeneously circularly polarized Gaussian beam are distributed around the spatial origin at $t = 19$ ps and are in the spin-up state [Fig.~\ref{fig:2}(a)].
By contrast, when the spins are excited using the vector vortex beam [Eq.~(\ref{eq:1}), with $l = 1$], a twofold symmetric structure of the spin state is observed at $t = 10$ ps, as shown in Fig.~\ref{fig:2}(b); in this structure, the spin-up and spin-down states are repeated in a circular pattern.
Since the vector vortex beam has a doughnut-shaped intensity profile in space originating from a topological singularity, spins are not excited into the spatial origin corresponding to the dark spot.
The observed spin texture with a twofold symmetric structure indicates the presence of the spin configuration shown in Fig.~\ref{fig:1}(c) and proves that the polarization pattern of the vector vortex beam is imprinted on the spatial structure of the electron spin states in the semiconductor.
Other spin textures can be easily generated using the vector vortex beam with different topological charges $l$, since the repetition rate of helicity varies with $l$, as shown in Eq.~(\ref{eq:1}).
Figure~\ref{fig:2}(c) shows the spin distribution at $t = 10$ ps excited by the vector vortex beam with $l = 2$.
The spatial period of the helicity of the excitation beam is half of that with $l = 1$.
Hence, the photoexcited spin texture has a fourfold symmetric structure, unlike the twofold symmetric spin texture excited by the vector vortex beam with $l = 1$.
In addition, the dark spot without spin excitation enlarges due to the increase in the topological charge.
Thus, the periodic helicity structure of the vector vortex beam enables the direct excitation of complex spin textures by a single beam.

Next, we investigate the spatiotemporal dynamics of the spin texture.
Figure~\ref{fig:3}(a) shows the spin distribution $s_z(x,y)$ at $t = 164$ ps under homogeneous spin excitation [Fig.~\ref{fig:2}(a)].
The observed spin texture exhibits a striped pattern, with repeated spin-up and spin-down states along the $x$ direction.
Figures~\ref{fig:3}(b) and \ref{fig:3}(c) show the time evolution of the spin distribution along the $x$ and $y$ directions, corresponding to $s_z(x,y=0,t)$ and  $s_z(x=0,y,t)$, respectively.
The locally excited electron spins diffuse in the QW plane from the initial spin distribution and precess with momentum-dependent SO effective magnetic fields with $t$.
The spins moving along the $x$ direction precess and form the helical spin pattern by the SO effective magnetic fields with time evolution, whereas the spins moving along the $y$ direction diffuse without spin precession.
Under the PSH condition in a (001) QW, where the Rashba and Dresselhaus SO interactions are balanced, the precession vector of electron spins is expressed as
\begin{eqnarray}
{\bf \Omega}=\frac{2k_{F}}{\hbar}
\begin{pmatrix}
0\\
(-\alpha+\tilde{\beta})\cos\theta
\end{pmatrix}
+{\bf \Omega}^{(3)}
\label{eq:2},
\end{eqnarray}
where $\hbar$ is the reduced Planck constant; $k_F=\sqrt{2\pi n}$ is the Fermi wave number; $\alpha$ is the Rashba coefficient; and $\tilde{\beta}=\beta_1-\beta_3$, where $\beta_1$ and $\beta_3$ are the linear and cubic Dresselhaus coefficients, respectively. $\theta$ is the angle between the wave vector and the $[1\bar{1}0]$ crystal axis.
The first term of Eq.~(\ref{eq:2}) means that uniaxially oriented effective magnetic fields with the SU(2) symmetry act on moving electron spins.
This leads to the formation of a PSH texture with helical spin stripes; this texture is robust to the spin-independent scattering.
By contrast, the third angular harmonics of ${\bf \Omega}^{(3)}=2k_F\beta_3(\sin3\theta{\bf e}_x-\cos3\theta{\bf e}_y)/\hbar$ destroys the PSH texture.
The observed spin stripe pattern suggests that the SO effective magnetic field induced by the studied QW structure is close to the PSH state.
The helical spin wavelength, which corresponds to the period of the spin stripe pattern, converges to the eigenvalue of the PSH state with $t$.
According to the residual helical spin mode at $t = 300$ ps, as shown in Fig.~\ref{fig:3}(b), the spin precession length $\lambda_\text{SO}$ is $11.0\pm0.3$ {\textmu}m. The SO parameter is obtained using $\lambda_\text{SO}$ as $-\alpha+\tilde{\beta}= \pi\hbar^2/m\lambda_\text{SO} = 3.2$ meV$\text{\AA}$, and the ratio $|\alpha /\tilde{\beta}|$ is estimated to be approximately 1.2\cite{note2022}.

The electron spins excited into the twofold symmetric spin texture by the vector vortex beam also diffuse in the QW plane and undergo the SO effective magnetic fields with $t$.
Figure~\ref{fig:3}(e) shows the spin distribution $s_z(x,y)$ at $t = 160$ ps, which is the time elapsed from the spin distribution in Fig.~\ref{fig:2}(b).
The distance between the excited spin-up states in Fig.~\ref{fig:2}(b) is intentionally set to $\lambda_\text{SO}$ in the PSH state.
At $t = 160$ ps, the dark spot of the spins at the spatial origin, where spins are not excited due to the doughnut-shaped light intensity profile, is filled by the inflow of electron spins.
Moreover, the spin texture becomes a striped pattern, as in the case of spin excitation by the homogeneously polarized beam.
Figures~\ref{fig:3}(f) and \ref{fig:3}(g) show the time evolution of the spin distributions $s_z(x,0,t)$ and $s_z(0,y,t)$ in the case of the vector vortex beam spin excitation.
The spin distribution in the $x$ direction broadens with $t$, forming a helical spin wave with the same $\lambda_\text{SO}$ in the case of spin excitation by the homogeneously polarized beam.
In the $y$ direction, the spin distribution broadens in the spin-down state without spin rotation because the SO field acting on the electron spins moving along the $y$ direction is canceled out.
As shown in Fig.~\ref{fig:3}(f), the helical spin wave converges to the spin wave with the intrinsic spin precession length of PSH, as is the case with the homogeneously polarized spin excitation, but its formation is faster than that in the latter case shown in Fig.~\ref{fig:3}(b).
The difference in the formation time of PSH modes between homogeneously polarized beam and vector vortex beam spin excitation is caused by the wave number distributions of the spin waves in the initial spin distribution.
Figures~\ref{fig:3}(d) and \ref{fig:3}(h) show the time evolution of the wave-mode distribution in the $x$ direction, analyzed by the fast Fourier transformation (FFT) of Figs.~\ref{fig:3}(b) and \ref{fig:3}(f), respectively.
The initial spin state with Gaussian distribution excited by the homogeneously polarized beam includes various spin wave components with different wave numbers.
Thus, the wave-mode distribution also has the Gaussian distribution centered at $q_x = 0$ in the wave number space, as shown in Fig.~\ref{fig:3}(d).
As the spin waves except the long-lived spin wave of the PSH mode of $q_\text{PSH} = 0.6$ {\textmu}m$^{-1}$ are relaxed, the helical spin wave with the eigenvalue of the spin precession length of the PSH state remains in real space over time.
The initial spin distribution has a periodic structure in the case of vector vortex beam excitation, unlike that in the case of homogeneous spin excitation.
When the distance between spin-up states is consistent with the spin precession length of the PSH, the dominant spin wave mode is already the PSH mode in the wave number space at the time of spin excitation, as shown in Fig.~\ref{fig:3}(h).
That is, the eigenwave of the PSH state is directly excited by the vector vortex beam.
Consequently, the helical spin wave of the PSH emerges in real space faster than that excited by the homogeneous polarized beam.
The relaxation times of the PSH mode extracted from the time evolution of the amplitude of the spin wave modes at $q_x = q_\text{PSH} = 0.6$ {\textmu}m$^{-1}$ of vector vortex beam and homogeneous beam excitation are $189\pm4$ and $158\pm5$ ps, respectively.
The difference is small because the dephasing of the PSH mode is determined by the balance between $\alpha$ and $\tilde{\beta}$ is independent of the polarization distribution of the excitation light.

\begin{figure}[b]
\includegraphics[scale=1]{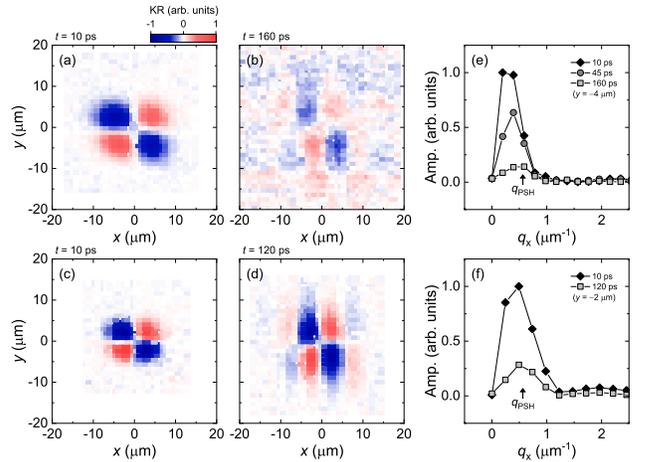}
\caption{\label{fig:4} Spatial spin distribution excited by $\pi/4$-rotated vector vortex beam with $l=1$.
$s_z(x,y)$ at (a) $t = 10$ ps and (b) $t = 160$ ps under large-spot spin excitation.
$s_z(x,y)$ at (c) $t = 10$ ps and (d) $t = 120$ ps under small-spot spin excitation.
Spin wave mode distribution in $x$ direction at different $t$ under (e) large-spot and (f) small-spot excitation.}
\end{figure}

As shown above, vector vortex beams can imprint their structured polarization distributions on electron spins in semiconductor QWs as spatial structures with specific wave modes.
However, for the efficient generation of the helical spin texture of the PSH, the excited spin mode in the periodic direction of the PSH stripe pattern ($x$ direction) should match the eigenmode of the spin wave in the PSH.
Figure~\ref{fig:4}(a) shows the spin distribution $s_z(x,y)$ at $t = 10$ ps excited by a $\pi/4$-rotated vector vortex beam with $l = 1$.
The excited spin texture is rotated by $\pi/4$ relative to that in Fig.~\ref{fig:2}(b), but its size is the same.
The periodic structure consists of spin-down and spin-up states along the positive $x$ direction, and its phase is inverted between positive and negative $y$ values.
The rotated spin texture is diluted and destroyed over time, as shown in Fig.~\ref{fig:4}(b), although the PSH texture is clearly observed in the case of the initial spin texture shown in Fig.~\ref{fig:2}(b).
For the $\pi/4$-rotated spin texture, because the distance between the spin-up and spin-down states parallel to the spin stripe direction becomes longer than half the PSH length, the excited spin mode does not match the eigenmode of the PSH texture, thereby disturbing the formation of the spin texture in real space.
Figure~\ref{fig:4}(c) shows the spin distribution $s_z(x,y)$ at $t = 10$ ps with a smaller spot size than that in Fig.~\ref{fig:4}(a).
The distance between the excited spin-up and spin-down states parallel to the $x$ axis is approximately 5 {\textmu}m, which is half the PSH length.
The spin distribution evolves with time into a spin texture with spin wave structures with opposite phases in positive and negative $y$, as shown in Fig.~\ref{fig:4}(d).
Figures~\ref{fig:4}(e) and \ref{fig:4}(f) show the wave-mode distributions at different $t$ values; these are FFT analysis findings from the spatial spin distributions under large-spot [Fig.~\ref{fig:4}(a)] and small-spot [Fig.~\ref{fig:4}(c)] excitation, respectively.
For large-spot spin excitation, the dominant spin wave mode at the time of photoexcitation ($t = 10$ ps) is not the eigenmode of the PSH $q_\text{PSH}$.
Subsequently, the initial spin wave mode decays, and the dominant wave modes shift toward the eigenvalue of the PSH with increasing time ($t = 45$ and 160 ps).
By contrast, for small-spot spin excitation, the dominant spin wave modes do not change with time relative to the initial spin distribution.
Hence, the spatial spin mode excited by the small-spot vector vortex beam is consistent with $q_\text{PSH}$.
Consequently, the decrease in the spin wave mode is suppressed, and two helical spin patterns with opposite phases are clearly observed in real space with electron spin diffusion.
Therefore, a new spin texture with two PSH wave modes having opposite phases can be produced by a single vector vortex beam tuning a proper helicity period and azimuthal angle with respect to the helical spin stripe by only adjusting the spot size and angle of the phase plate; the use of a conventional homogeneously polarized beam cannot accomplish this process.

In conclusion, we directly generate electron spin textures by imprinting the polarization distribution of topologically structured light.
The vector vortex beam imprints its periodic helicity structure on the spatial pattern of electron spin states, where the spin-up and spin-down states are repeated in a circular pattern.
The excitation of the periodic spin texture by the vector vortex beam corresponds to the excitation of a specific spatial spin wave mode.
Matching the excited spin mode with the eigenmode of the long-lived spin wave of the PSH, the formation of the PSH texture in real space is accelerated because the extra spin wave modes that prevent the formation of the PSH texture are not excited.
We also show that a discriminative spin texture, such as the presented PSH wave modes with opposite phases, can be produced even from a single beam by using a combination of helicity patterns of structured light and momentum-dependent SO effective magnetic fields.
Our findings are valuable for the preparation and control of arbitrary spin textures in semiconductors and are of great importance for high-order quantum media conversion and spintronic applications using spin textures.

This work is supported by JSPS KAKENHI Grant Nos. JP18K14113, JP21K14528, JP21H04647, and JP22H01981, the Shimadzu Science Foundation, the Murata Science Foundation, JST FOREST Program Grant No. JPMJFR203C, and the Cooperative Research Project Program of RIEC and Center for Spintronics Research Network, Tohoku University.

\nocite{*}

\bibliography{Ishihara_etal_VBtoST_v6}

\begin{thebibliography}{33}%
\makeatletter
\providecommand \@ifxundefined [1]{%
 \@ifx{#1\undefined}
}%
\providecommand \@ifnum [1]{%
 \ifnum #1\expandafter \@firstoftwo
 \else \expandafter \@secondoftwo
 \fi
}%
\providecommand \@ifx [1]{%
 \ifx #1\expandafter \@firstoftwo
 \else \expandafter \@secondoftwo
 \fi
}%
\providecommand \natexlab [1]{#1}%
\providecommand \enquote  [1]{``#1''}%
\providecommand \bibnamefont  [1]{#1}%
\providecommand \bibfnamefont [1]{#1}%
\providecommand \citenamefont [1]{#1}%
\providecommand \href@noop [0]{\@secondoftwo}%
\providecommand \href [0]{\begingroup \@sanitize@url \@href}%
\providecommand \@href[1]{\@@startlink{#1}\@@href}%
\providecommand \@@href[1]{\endgroup#1\@@endlink}%
\providecommand \@sanitize@url [0]{\catcode `\\12\catcode `\$12\catcode
  `\&12\catcode `\#12\catcode `\^12\catcode `\_12\catcode `\%12\relax}%
\providecommand \@@startlink[1]{}%
\providecommand \@@endlink[0]{}%
\providecommand \url  [0]{\begingroup\@sanitize@url \@url }%
\providecommand \@url [1]{\endgroup\@href {#1}{\urlprefix }}%
\providecommand \urlprefix  [0]{URL }%
\providecommand \Eprint [0]{\href }%
\providecommand \doibase [0]{https://doi.org/}%
\providecommand \selectlanguage [0]{\@gobble}%
\providecommand \bibinfo  [0]{\@secondoftwo}%
\providecommand \bibfield  [0]{\@secondoftwo}%
\providecommand \translation [1]{[#1]}%
\providecommand \BibitemOpen [0]{}%
\providecommand \bibitemStop [0]{}%
\providecommand \bibitemNoStop [0]{.\EOS\space}%
\providecommand \EOS [0]{\spacefactor3000\relax}%
\providecommand \BibitemShut  [1]{\csname bibitem#1\endcsname}%
\let\auto@bib@innerbib\@empty
\bibitem [{\citenamefont {Hell}\ and\ \citenamefont
  {Wichmann}(1994)}]{Hell1994}%
  \BibitemOpen
  \bibfield  {author} {\bibinfo {author} {\bibfnamefont {S.~W.}\ \bibnamefont
  {Hell}}\ and\ \bibinfo {author} {\bibfnamefont {J.}~\bibnamefont
  {Wichmann}},\ }\href {https://doi.org/10.1364/OL.19.000780} {\bibfield
  {journal} {\bibinfo  {journal} {Opt. Lett.}\ }\textbf {\bibinfo {volume}
  {19}},\ \bibinfo {pages} {780} (\bibinfo {year} {1994})}\BibitemShut
  {NoStop}%
\bibitem [{\citenamefont {Willig}\ \emph {et~al.}(2007)\citenamefont {Willig},
  \citenamefont {Harke}, \citenamefont {Medda},\ and\ \citenamefont
  {Hell}}]{Willing2007}%
  \BibitemOpen
  \bibfield  {author} {\bibinfo {author} {\bibfnamefont {K.~I.}\ \bibnamefont
  {Willig}}, \bibinfo {author} {\bibfnamefont {B.}~\bibnamefont {Harke}},
  \bibinfo {author} {\bibfnamefont {R.}~\bibnamefont {Medda}},\ and\ \bibinfo
  {author} {\bibfnamefont {S.~W.}\ \bibnamefont {Hell}},\ }\href
  {https://doi.org/10.1038/nmeth1108} {\bibfield  {journal} {\bibinfo
  {journal} {Nat. Methods}\ }\textbf {\bibinfo {volume} {4}},\ \bibinfo {pages}
  {915} (\bibinfo {year} {2007})}\BibitemShut {NoStop}%
\bibitem [{\citenamefont {Toyoda}\ \emph {et~al.}(2012)\citenamefont {Toyoda},
  \citenamefont {Miyamoto}, \citenamefont {Aoki}, \citenamefont {Morita},\ and\
  \citenamefont {Omatsu}}]{Toyoda2012}%
  \BibitemOpen
  \bibfield  {author} {\bibinfo {author} {\bibfnamefont {K.}~\bibnamefont
  {Toyoda}}, \bibinfo {author} {\bibfnamefont {K.}~\bibnamefont {Miyamoto}},
  \bibinfo {author} {\bibfnamefont {N.}~\bibnamefont {Aoki}}, \bibinfo {author}
  {\bibfnamefont {R.}~\bibnamefont {Morita}},\ and\ \bibinfo {author}
  {\bibfnamefont {T.}~\bibnamefont {Omatsu}},\ }\href
  {https://doi.org/10.1021/nl301347j} {\bibfield  {journal} {\bibinfo
  {journal} {Nano Lett.}\ }\textbf {\bibinfo {volume} {12}},\ \bibinfo {pages}
  {3645} (\bibinfo {year} {2012})}\BibitemShut {NoStop}%
\bibitem [{\citenamefont {Toyoda}\ \emph {et~al.}(2013)\citenamefont {Toyoda},
  \citenamefont {Takahashi}, \citenamefont {Takizawa}, \citenamefont
  {Tokizane}, \citenamefont {Miyamoto}, \citenamefont {Morita},\ and\
  \citenamefont {Omatsu}}]{Toyoda2013}%
  \BibitemOpen
  \bibfield  {author} {\bibinfo {author} {\bibfnamefont {K.}~\bibnamefont
  {Toyoda}}, \bibinfo {author} {\bibfnamefont {F.}~\bibnamefont {Takahashi}},
  \bibinfo {author} {\bibfnamefont {S.}~\bibnamefont {Takizawa}}, \bibinfo
  {author} {\bibfnamefont {Y.}~\bibnamefont {Tokizane}}, \bibinfo {author}
  {\bibfnamefont {K.}~\bibnamefont {Miyamoto}}, \bibinfo {author}
  {\bibfnamefont {R.}~\bibnamefont {Morita}},\ and\ \bibinfo {author}
  {\bibfnamefont {T.}~\bibnamefont {Omatsu}},\ }\href
  {https://doi.org/10.1103/PhysRevLett.110.143603} {\bibfield  {journal}
  {\bibinfo  {journal} {Phys. Rev. Lett.}\ }\textbf {\bibinfo {volume} {110}},\
  \bibinfo {pages} {143603} (\bibinfo {year} {2013})}\BibitemShut {NoStop}%
\bibitem [{\citenamefont {O'Neil}\ \emph {et~al.}(2002)\citenamefont {O'Neil},
  \citenamefont {MacVicar}, \citenamefont {Allen},\ and\ \citenamefont
  {Padgett}}]{ONeil2002}%
  \BibitemOpen
  \bibfield  {author} {\bibinfo {author} {\bibfnamefont {A.~T.}\ \bibnamefont
  {O'Neil}}, \bibinfo {author} {\bibfnamefont {I.}~\bibnamefont {MacVicar}},
  \bibinfo {author} {\bibfnamefont {L.}~\bibnamefont {Allen}},\ and\ \bibinfo
  {author} {\bibfnamefont {M.~J.}\ \bibnamefont {Padgett}},\ }\href
  {https://doi.org/10.1103/PhysRevLett.88.053601} {\bibfield  {journal}
  {\bibinfo  {journal} {Phys. Rev. Lett.}\ }\textbf {\bibinfo {volume} {88}},\
  \bibinfo {pages} {053601} (\bibinfo {year} {2002})}\BibitemShut {NoStop}%
\bibitem [{\citenamefont {Curtis}\ and\ \citenamefont
  {Grier}(2003)}]{Curtis2003}%
  \BibitemOpen
  \bibfield  {author} {\bibinfo {author} {\bibfnamefont {J.~E.}\ \bibnamefont
  {Curtis}}\ and\ \bibinfo {author} {\bibfnamefont {D.~G.}\ \bibnamefont
  {Grier}},\ }\href {https://doi.org/10.1103/PhysRevLett.90.133901} {\bibfield
  {journal} {\bibinfo  {journal} {Phys. Rev. Lett.}\ }\textbf {\bibinfo
  {volume} {90}},\ \bibinfo {pages} {133901} (\bibinfo {year}
  {2003})}\BibitemShut {NoStop}%
\bibitem [{\citenamefont {Ndagano}\ \emph {et~al.}(2018)\citenamefont
  {Ndagano}, \citenamefont {Nape}, \citenamefont {Cox}, \citenamefont
  {Rosales-Guzman},\ and\ \citenamefont {Forbes}}]{Ndagano2018}%
  \BibitemOpen
  \bibfield  {author} {\bibinfo {author} {\bibfnamefont {B.}~\bibnamefont
  {Ndagano}}, \bibinfo {author} {\bibfnamefont {I.}~\bibnamefont {Nape}},
  \bibinfo {author} {\bibfnamefont {M.~A.}\ \bibnamefont {Cox}}, \bibinfo
  {author} {\bibfnamefont {C.}~\bibnamefont {Rosales-Guzman}},\ and\ \bibinfo
  {author} {\bibfnamefont {A.}~\bibnamefont {Forbes}},\ }\href
  {https://doi.org/10.1109/JLT.2017.2766760} {\bibfield  {journal} {\bibinfo
  {journal} {J. Light. Technol.}\ }\textbf {\bibinfo {volume} {36}},\ \bibinfo
  {pages} {292} (\bibinfo {year} {2018})}\BibitemShut {NoStop}%
\bibitem [{\citenamefont {Zhao}\ and\ \citenamefont {Wang}(2015)}]{Zhao2015}%
  \BibitemOpen
  \bibfield  {author} {\bibinfo {author} {\bibfnamefont {Y.}~\bibnamefont
  {Zhao}}\ and\ \bibinfo {author} {\bibfnamefont {J.}~\bibnamefont {Wang}},\
  }\href {https://doi.org/10.1364/OL.40.004843} {\bibfield  {journal} {\bibinfo
   {journal} {Opt. Lett.}\ }\textbf {\bibinfo {volume} {40}},\ \bibinfo {pages}
  {4843} (\bibinfo {year} {2015})}\BibitemShut {NoStop}%
\bibitem [{\citenamefont {Milione}\ \emph {et~al.}(2015)\citenamefont
  {Milione}, \citenamefont {Lavery}, \citenamefont {Huang}, \citenamefont
  {Ren}, \citenamefont {Xie}, \citenamefont {Nguyen}, \citenamefont {Karimi},
  \citenamefont {Marrucci}, \citenamefont {Nolan}, \citenamefont {Alfano},\
  and\ \citenamefont {Willner}}]{Milione2015}%
  \BibitemOpen
  \bibfield  {author} {\bibinfo {author} {\bibfnamefont {G.}~\bibnamefont
  {Milione}}, \bibinfo {author} {\bibfnamefont {M.~P.~J.}\ \bibnamefont
  {Lavery}}, \bibinfo {author} {\bibfnamefont {H.}~\bibnamefont {Huang}},
  \bibinfo {author} {\bibfnamefont {Y.}~\bibnamefont {Ren}}, \bibinfo {author}
  {\bibfnamefont {G.}~\bibnamefont {Xie}}, \bibinfo {author} {\bibfnamefont
  {T.~A.}\ \bibnamefont {Nguyen}}, \bibinfo {author} {\bibfnamefont
  {E.}~\bibnamefont {Karimi}}, \bibinfo {author} {\bibfnamefont
  {L.}~\bibnamefont {Marrucci}}, \bibinfo {author} {\bibfnamefont {D.~A.}\
  \bibnamefont {Nolan}}, \bibinfo {author} {\bibfnamefont {R.~R.}\ \bibnamefont
  {Alfano}},\ and\ \bibinfo {author} {\bibfnamefont {A.~E.}\ \bibnamefont
  {Willner}},\ }\href {https://doi.org/10.1364/OL.40.001980} {\bibfield
  {journal} {\bibinfo  {journal} {Opt. Lett.}\ }\textbf {\bibinfo {volume}
  {40}},\ \bibinfo {pages} {1980} (\bibinfo {year} {2015})}\BibitemShut
  {NoStop}%
\bibitem [{\citenamefont {Berg-Johansen}\ \emph {et~al.}(2015)\citenamefont
  {Berg-Johansen}, \citenamefont {T\"{o}ppel}, \citenamefont {Stiller},
  \citenamefont {Banzer}, \citenamefont {Ornigotti}, \citenamefont {Giacobino},
  \citenamefont {Leuchs}, \citenamefont {Aiello},\ and\ \citenamefont
  {Marquardt}}]{Berg-Johansen2015}%
  \BibitemOpen
  \bibfield  {author} {\bibinfo {author} {\bibfnamefont {S.}~\bibnamefont
  {Berg-Johansen}}, \bibinfo {author} {\bibfnamefont {F.}~\bibnamefont
  {T\"{o}ppel}}, \bibinfo {author} {\bibfnamefont {B.}~\bibnamefont {Stiller}},
  \bibinfo {author} {\bibfnamefont {P.}~\bibnamefont {Banzer}}, \bibinfo
  {author} {\bibfnamefont {M.}~\bibnamefont {Ornigotti}}, \bibinfo {author}
  {\bibfnamefont {E.}~\bibnamefont {Giacobino}}, \bibinfo {author}
  {\bibfnamefont {G.}~\bibnamefont {Leuchs}}, \bibinfo {author} {\bibfnamefont
  {A.}~\bibnamefont {Aiello}},\ and\ \bibinfo {author} {\bibfnamefont
  {C.}~\bibnamefont {Marquardt}},\ }\href
  {https://doi.org/10.1364/OPTICA.2.000864} {\bibfield  {journal} {\bibinfo
  {journal} {Optica}\ }\textbf {\bibinfo {volume} {2}},\ \bibinfo {pages} {864}
  (\bibinfo {year} {2015})}\BibitemShut {NoStop}%
\bibitem [{\citenamefont {Hu}\ \emph {et~al.}(2019)\citenamefont {Hu},
  \citenamefont {Zhao}, \citenamefont {Zhu}, \citenamefont {Gao},\ and\
  \citenamefont {Rosales-Guzm\'{a}n}}]{Hu2019}%
  \BibitemOpen
  \bibfield  {author} {\bibinfo {author} {\bibfnamefont {X.-B.}\ \bibnamefont
  {Hu}}, \bibinfo {author} {\bibfnamefont {B.}~\bibnamefont {Zhao}}, \bibinfo
  {author} {\bibfnamefont {Z.-H.}\ \bibnamefont {Zhu}}, \bibinfo {author}
  {\bibfnamefont {W.}~\bibnamefont {Gao}},\ and\ \bibinfo {author}
  {\bibfnamefont {C.}~\bibnamefont {Rosales-Guzm\'{a}n}},\ }\href
  {https://doi.org/10.1364/OL.44.003070} {\bibfield  {journal} {\bibinfo
  {journal} {Opt. Lett.}\ }\textbf {\bibinfo {volume} {44}},\ \bibinfo {pages}
  {3070} (\bibinfo {year} {2019})}\BibitemShut {NoStop}%
\bibitem [{\citenamefont {Rashba}(1960)}]{Rashba1960}%
  \BibitemOpen
  \bibfield  {author} {\bibinfo {author} {\bibfnamefont {E.~I.}\ \bibnamefont
  {Rashba}},\ }\href@noop {} {\bibfield  {journal} {\bibinfo  {journal}
  {Sov.Phys. Solid State}\ }\textbf {\bibinfo {volume} {2}},\ \bibinfo {pages}
  {1109} (\bibinfo {year} {1960})}\BibitemShut {NoStop}%
\bibitem [{\citenamefont {Dresselhaus}(1955)}]{Dresselhaus1955}%
  \BibitemOpen
  \bibfield  {author} {\bibinfo {author} {\bibfnamefont {G.}~\bibnamefont
  {Dresselhaus}},\ }\href {https://doi.org/10.1103/PhysRev.100.580} {\bibfield
  {journal} {\bibinfo  {journal} {Phys. Rev.}\ }\textbf {\bibinfo {volume}
  {100}},\ \bibinfo {pages} {580} (\bibinfo {year} {1955})}\BibitemShut
  {NoStop}%
\bibitem [{\citenamefont {Schliemann}\ and\ \citenamefont
  {Loss}(2003)}]{Schliemann2003}%
  \BibitemOpen
  \bibfield  {author} {\bibinfo {author} {\bibfnamefont {J.}~\bibnamefont
  {Schliemann}}\ and\ \bibinfo {author} {\bibfnamefont {D.}~\bibnamefont
  {Loss}},\ }\href {https://doi.org/10.1103/PhysRevB.68.165311} {\bibfield
  {journal} {\bibinfo  {journal} {Phys. Rev. B}\ }\textbf {\bibinfo {volume}
  {68}},\ \bibinfo {pages} {165311} (\bibinfo {year} {2003})}\BibitemShut
  {NoStop}%
\bibitem [{\citenamefont {Bernevig}\ \emph {et~al.}(2006)\citenamefont
  {Bernevig}, \citenamefont {Orenstein},\ and\ \citenamefont
  {Zhang}}]{Bernevig2006}%
  \BibitemOpen
  \bibfield  {author} {\bibinfo {author} {\bibfnamefont {B.~A.}\ \bibnamefont
  {Bernevig}}, \bibinfo {author} {\bibfnamefont {J.}~\bibnamefont
  {Orenstein}},\ and\ \bibinfo {author} {\bibfnamefont {S.~C.}\ \bibnamefont
  {Zhang}},\ }\href {https://doi.org/10.1103/PhysRevLett.97.236601} {\bibfield
  {journal} {\bibinfo  {journal} {Phys. Rev. Lett.}\ }\textbf {\bibinfo
  {volume} {97}},\ \bibinfo {pages} {236601} (\bibinfo {year}
  {2006})}\BibitemShut {NoStop}%
\bibitem [{\citenamefont {Koralek}\ \emph {et~al.}(2009)\citenamefont
  {Koralek}, \citenamefont {Weber}, \citenamefont {Orenstein}, \citenamefont
  {Bernevig}, \citenamefont {Zhang}, \citenamefont {Mack},\ and\ \citenamefont
  {Awschalom}}]{Koralek2009}%
  \BibitemOpen
  \bibfield  {author} {\bibinfo {author} {\bibfnamefont {J.~D.}\ \bibnamefont
  {Koralek}}, \bibinfo {author} {\bibfnamefont {C.~P.}\ \bibnamefont {Weber}},
  \bibinfo {author} {\bibfnamefont {J.}~\bibnamefont {Orenstein}}, \bibinfo
  {author} {\bibfnamefont {B.~a.}\ \bibnamefont {Bernevig}}, \bibinfo {author}
  {\bibfnamefont {S.-C.}\ \bibnamefont {Zhang}}, \bibinfo {author}
  {\bibfnamefont {S.}~\bibnamefont {Mack}},\ and\ \bibinfo {author}
  {\bibfnamefont {D.~D.}\ \bibnamefont {Awschalom}},\ }\href
  {https://doi.org/10.1038/nature07871} {\bibfield  {journal} {\bibinfo
  {journal} {Nature}\ }\textbf {\bibinfo {volume} {458}},\ \bibinfo {pages}
  {610} (\bibinfo {year} {2009})}\BibitemShut {NoStop}%
\bibitem [{\citenamefont {Walser}\ \emph {et~al.}(2012)\citenamefont {Walser},
  \citenamefont {Reichl}, \citenamefont {Wegscheider},\ and\ \citenamefont
  {Salis}}]{Walser2012}%
  \BibitemOpen
  \bibfield  {author} {\bibinfo {author} {\bibfnamefont {M.~P.}\ \bibnamefont
  {Walser}}, \bibinfo {author} {\bibfnamefont {C.}~\bibnamefont {Reichl}},
  \bibinfo {author} {\bibfnamefont {W.}~\bibnamefont {Wegscheider}},\ and\
  \bibinfo {author} {\bibfnamefont {G.}~\bibnamefont {Salis}},\ }\href
  {https://doi.org/10.1038/nphys2383} {\bibfield  {journal} {\bibinfo
  {journal} {Nat. Phys.}\ }\textbf {\bibinfo {volume} {8}},\ \bibinfo {pages}
  {757} (\bibinfo {year} {2012})}\BibitemShut {NoStop}%
\bibitem [{\citenamefont {Ishihara}\ \emph {et~al.}(2013)\citenamefont
  {Ishihara}, \citenamefont {Ohno},\ and\ \citenamefont {Ohno}}]{Ishihara2013}%
  \BibitemOpen
  \bibfield  {author} {\bibinfo {author} {\bibfnamefont {J.}~\bibnamefont
  {Ishihara}}, \bibinfo {author} {\bibfnamefont {Y.}~\bibnamefont {Ohno}},\
  and\ \bibinfo {author} {\bibfnamefont {H.}~\bibnamefont {Ohno}},\ }\href
  {https://doi.org/10.7567/apex.7.013001} {\bibfield  {journal} {\bibinfo
  {journal} {Appl. Phys. Express}\ }\textbf {\bibinfo {volume} {7}},\ \bibinfo
  {pages} {013001} (\bibinfo {year} {2013})}\BibitemShut {NoStop}%
\bibitem [{\citenamefont {Ishihara}\ \emph {et~al.}(2014)\citenamefont
  {Ishihara}, \citenamefont {Ohno},\ and\ \citenamefont {Ohno}}]{Ishihara2014}%
  \BibitemOpen
  \bibfield  {author} {\bibinfo {author} {\bibfnamefont {J.}~\bibnamefont
  {Ishihara}}, \bibinfo {author} {\bibfnamefont {Y.}~\bibnamefont {Ohno}},\
  and\ \bibinfo {author} {\bibfnamefont {H.}~\bibnamefont {Ohno}},\ }\href
  {https://doi.org/10.7567/jjap.53.04em04} {\bibfield  {journal} {\bibinfo
  {journal} {Jpn. J. Appl. Phys.}\ }\textbf {\bibinfo {volume} {53}},\ \bibinfo
  {pages} {04EM04} (\bibinfo {year} {2014})}\BibitemShut {NoStop}%
\bibitem [{\citenamefont {Altmann}\ \emph {et~al.}(2014)\citenamefont
  {Altmann}, \citenamefont {Walser}, \citenamefont {Reichl}, \citenamefont
  {Wegscheider},\ and\ \citenamefont {Salis}}]{Altmann2014}%
  \BibitemOpen
  \bibfield  {author} {\bibinfo {author} {\bibfnamefont {P.}~\bibnamefont
  {Altmann}}, \bibinfo {author} {\bibfnamefont {M.~P.}\ \bibnamefont {Walser}},
  \bibinfo {author} {\bibfnamefont {C.}~\bibnamefont {Reichl}}, \bibinfo
  {author} {\bibfnamefont {W.}~\bibnamefont {Wegscheider}},\ and\ \bibinfo
  {author} {\bibfnamefont {G.}~\bibnamefont {Salis}},\ }\href
  {https://doi.org/10.1103/PhysRevB.90.201306} {\bibfield  {journal} {\bibinfo
  {journal} {Phys. Rev. B}\ }\textbf {\bibinfo {volume} {90}},\ \bibinfo
  {pages} {201306} (\bibinfo {year} {2014})}\BibitemShut {NoStop}%
\bibitem [{\citenamefont {Salis}\ \emph {et~al.}(2014)\citenamefont {Salis},
  \citenamefont {Walser}, \citenamefont {Altmann}, \citenamefont {Reichl},\
  and\ \citenamefont {Wegscheider}}]{Salis2014}%
  \BibitemOpen
  \bibfield  {author} {\bibinfo {author} {\bibfnamefont {G.}~\bibnamefont
  {Salis}}, \bibinfo {author} {\bibfnamefont {M.~P.}\ \bibnamefont {Walser}},
  \bibinfo {author} {\bibfnamefont {P.}~\bibnamefont {Altmann}}, \bibinfo
  {author} {\bibfnamefont {C.}~\bibnamefont {Reichl}},\ and\ \bibinfo {author}
  {\bibfnamefont {W.}~\bibnamefont {Wegscheider}},\ }\href
  {https://doi.org/10.1103/PhysRevB.89.045304} {\bibfield  {journal} {\bibinfo
  {journal} {Phys. Rev. B}\ }\textbf {\bibinfo {volume} {89}},\ \bibinfo
  {pages} {045304} (\bibinfo {year} {2014})}\BibitemShut {NoStop}%
\bibitem [{\citenamefont {Ishihara}\ \emph {et~al.}(2022)\citenamefont
  {Ishihara}, \citenamefont {Suzuki}, \citenamefont {Kitazawa}, \citenamefont
  {Mori}, \citenamefont {Ohno},\ and\ \citenamefont {Miyajima}}]{Ishihara2022}%
  \BibitemOpen
  \bibfield  {author} {\bibinfo {author} {\bibfnamefont {J.}~\bibnamefont
  {Ishihara}}, \bibinfo {author} {\bibfnamefont {T.}~\bibnamefont {Suzuki}},
  \bibinfo {author} {\bibfnamefont {G.}~\bibnamefont {Kitazawa}}, \bibinfo
  {author} {\bibfnamefont {T.}~\bibnamefont {Mori}}, \bibinfo {author}
  {\bibfnamefont {Y.}~\bibnamefont {Ohno}},\ and\ \bibinfo {author}
  {\bibfnamefont {K.}~\bibnamefont {Miyajima}},\ }\href
  {https://doi.org/10.1103/PhysRevB.105.144412} {\bibfield  {journal} {\bibinfo
   {journal} {Phys. Rev. B}\ }\textbf {\bibinfo {volume} {105}},\ \bibinfo
  {pages} {144412} (\bibinfo {year} {2022})}\BibitemShut {NoStop}%
\bibitem [{\citenamefont {Kohda}\ \emph {et~al.}(2012)\citenamefont {Kohda},
  \citenamefont {Lechner}, \citenamefont {Kunihashi}, \citenamefont
  {Dollinger}, \citenamefont {Olbrich}, \citenamefont {Sch\"onhuber},
  \citenamefont {Caspers}, \citenamefont {Bel'kov}, \citenamefont {Golub},
  \citenamefont {Weiss}, \citenamefont {Richter}, \citenamefont {Nitta},\ and\
  \citenamefont {Ganichev}}]{Kohda2012}%
  \BibitemOpen
  \bibfield  {author} {\bibinfo {author} {\bibfnamefont {M.}~\bibnamefont
  {Kohda}}, \bibinfo {author} {\bibfnamefont {V.}~\bibnamefont {Lechner}},
  \bibinfo {author} {\bibfnamefont {Y.}~\bibnamefont {Kunihashi}}, \bibinfo
  {author} {\bibfnamefont {T.}~\bibnamefont {Dollinger}}, \bibinfo {author}
  {\bibfnamefont {P.}~\bibnamefont {Olbrich}}, \bibinfo {author} {\bibfnamefont
  {C.}~\bibnamefont {Sch\"onhuber}}, \bibinfo {author} {\bibfnamefont
  {I.}~\bibnamefont {Caspers}}, \bibinfo {author} {\bibfnamefont {V.~V.}\
  \bibnamefont {Bel'kov}}, \bibinfo {author} {\bibfnamefont {L.~E.}\
  \bibnamefont {Golub}}, \bibinfo {author} {\bibfnamefont {D.}~\bibnamefont
  {Weiss}}, \bibinfo {author} {\bibfnamefont {K.}~\bibnamefont {Richter}},
  \bibinfo {author} {\bibfnamefont {J.}~\bibnamefont {Nitta}},\ and\ \bibinfo
  {author} {\bibfnamefont {S.~D.}\ \bibnamefont {Ganichev}},\ }\href
  {https://doi.org/10.1103/PhysRevB.86.081306} {\bibfield  {journal} {\bibinfo
  {journal} {Phys. Rev. B}\ }\textbf {\bibinfo {volume} {86}},\ \bibinfo
  {pages} {081306} (\bibinfo {year} {2012})}\BibitemShut {NoStop}%
\bibitem [{\citenamefont {Kunihashi}\ \emph {et~al.}(2016)\citenamefont
  {Kunihashi}, \citenamefont {Sanada}, \citenamefont {Gotoh}, \citenamefont
  {Onomitsu},\ and\ \citenamefont {Sogawa}}]{Kunihashi2016}%
  \BibitemOpen
  \bibfield  {author} {\bibinfo {author} {\bibfnamefont {Y.}~\bibnamefont
  {Kunihashi}}, \bibinfo {author} {\bibfnamefont {H.}~\bibnamefont {Sanada}},
  \bibinfo {author} {\bibfnamefont {H.}~\bibnamefont {Gotoh}}, \bibinfo
  {author} {\bibfnamefont {K.}~\bibnamefont {Onomitsu}},\ and\ \bibinfo
  {author} {\bibfnamefont {T.}~\bibnamefont {Sogawa}},\ }\href
  {https://doi.org/10.1038/ncomms10722} {\bibfield  {journal} {\bibinfo
  {journal} {Nat. Commun.}\ }\textbf {\bibinfo {volume} {14}},\ \bibinfo
  {pages} {10722} (\bibinfo {year} {2016})}\BibitemShut {NoStop}%
\bibitem [{\citenamefont {Altmann}\ \emph {et~al.}(2016)\citenamefont
  {Altmann}, \citenamefont {Hernandez}, \citenamefont {Ferreira}, \citenamefont
  {Kohda}, \citenamefont {Reichl}, \citenamefont {Wegscheider},\ and\
  \citenamefont {Salis}}]{Altmann2016}%
  \BibitemOpen
  \bibfield  {author} {\bibinfo {author} {\bibfnamefont {P.}~\bibnamefont
  {Altmann}}, \bibinfo {author} {\bibfnamefont {F.~G.~G.}\ \bibnamefont
  {Hernandez}}, \bibinfo {author} {\bibfnamefont {G.~J.}\ \bibnamefont
  {Ferreira}}, \bibinfo {author} {\bibfnamefont {M.}~\bibnamefont {Kohda}},
  \bibinfo {author} {\bibfnamefont {C.}~\bibnamefont {Reichl}}, \bibinfo
  {author} {\bibfnamefont {W.}~\bibnamefont {Wegscheider}},\ and\ \bibinfo
  {author} {\bibfnamefont {G.}~\bibnamefont {Salis}},\ }\href
  {https://doi.org/10.1103/PhysRevLett.116.196802} {\bibfield  {journal}
  {\bibinfo  {journal} {Phys. Rev. Lett.}\ }\textbf {\bibinfo {volume} {116}},\
  \bibinfo {pages} {196802} (\bibinfo {year} {2016})}\BibitemShut {NoStop}%
\bibitem [{\citenamefont {Passmann}\ \emph {et~al.}(2018)\citenamefont
  {Passmann}, \citenamefont {Anghel}, \citenamefont {Tischler}, \citenamefont
  {Poshakinskiy}, \citenamefont {Tarasenko}, \citenamefont {Karczewski},
  \citenamefont {Wojtowicz}, \citenamefont {Bristow},\ and\ \citenamefont
  {Betz}}]{Passmann2018}%
  \BibitemOpen
  \bibfield  {author} {\bibinfo {author} {\bibfnamefont {F.}~\bibnamefont
  {Passmann}}, \bibinfo {author} {\bibfnamefont {S.}~\bibnamefont {Anghel}},
  \bibinfo {author} {\bibfnamefont {T.}~\bibnamefont {Tischler}}, \bibinfo
  {author} {\bibfnamefont {A.~V.}\ \bibnamefont {Poshakinskiy}}, \bibinfo
  {author} {\bibfnamefont {S.~A.}\ \bibnamefont {Tarasenko}}, \bibinfo {author}
  {\bibfnamefont {G.}~\bibnamefont {Karczewski}}, \bibinfo {author}
  {\bibfnamefont {T.}~\bibnamefont {Wojtowicz}}, \bibinfo {author}
  {\bibfnamefont {A.~D.}\ \bibnamefont {Bristow}},\ and\ \bibinfo {author}
  {\bibfnamefont {M.}~\bibnamefont {Betz}},\ }\href
  {https://doi.org/10.1103/PhysRevB.97.201413} {\bibfield  {journal} {\bibinfo
  {journal} {Phys. Rev. B}\ }\textbf {\bibinfo {volume} {97}},\ \bibinfo
  {pages} {201413} (\bibinfo {year} {2018})}\BibitemShut {NoStop}%
\bibitem [{\citenamefont {Anghel}\ \emph {et~al.}(2018)\citenamefont {Anghel},
  \citenamefont {Passmann}, \citenamefont {Singh}, \citenamefont {Ruppert},
  \citenamefont {Poshakinskiy}, \citenamefont {Tarasenko}, \citenamefont
  {Moore}, \citenamefont {Yusa}, \citenamefont {Mano}, \citenamefont {Noda},
  \citenamefont {Li}, \citenamefont {Bristow},\ and\ \citenamefont
  {Betz}}]{Anghel2018}%
  \BibitemOpen
  \bibfield  {author} {\bibinfo {author} {\bibfnamefont {S.}~\bibnamefont
  {Anghel}}, \bibinfo {author} {\bibfnamefont {F.}~\bibnamefont {Passmann}},
  \bibinfo {author} {\bibfnamefont {A.}~\bibnamefont {Singh}}, \bibinfo
  {author} {\bibfnamefont {C.}~\bibnamefont {Ruppert}}, \bibinfo {author}
  {\bibfnamefont {A.~V.}\ \bibnamefont {Poshakinskiy}}, \bibinfo {author}
  {\bibfnamefont {S.~A.}\ \bibnamefont {Tarasenko}}, \bibinfo {author}
  {\bibfnamefont {J.~N.}\ \bibnamefont {Moore}}, \bibinfo {author}
  {\bibfnamefont {G.}~\bibnamefont {Yusa}}, \bibinfo {author} {\bibfnamefont
  {T.}~\bibnamefont {Mano}}, \bibinfo {author} {\bibfnamefont {T.}~\bibnamefont
  {Noda}}, \bibinfo {author} {\bibfnamefont {X.}~\bibnamefont {Li}}, \bibinfo
  {author} {\bibfnamefont {A.~D.}\ \bibnamefont {Bristow}},\ and\ \bibinfo
  {author} {\bibfnamefont {M.}~\bibnamefont {Betz}},\ }\href
  {https://doi.org/10.1103/PhysRevB.97.125410} {\bibfield  {journal} {\bibinfo
  {journal} {Phys. Rev. B}\ }\textbf {\bibinfo {volume} {97}},\ \bibinfo
  {pages} {125410} (\bibinfo {year} {2018})}\BibitemShut {NoStop}%
\bibitem [{\citenamefont {Anghel}\ \emph {et~al.}(2020)\citenamefont {Anghel},
  \citenamefont {Passmann}, \citenamefont {Schiller}, \citenamefont {Moore},
  \citenamefont {Yusa}, \citenamefont {Mano}, \citenamefont {Noda},
  \citenamefont {Betz},\ and\ \citenamefont {Bristow}}]{Anghel2020}%
  \BibitemOpen
  \bibfield  {author} {\bibinfo {author} {\bibfnamefont {S.}~\bibnamefont
  {Anghel}}, \bibinfo {author} {\bibfnamefont {F.}~\bibnamefont {Passmann}},
  \bibinfo {author} {\bibfnamefont {K.~J.}\ \bibnamefont {Schiller}}, \bibinfo
  {author} {\bibfnamefont {J.~N.}\ \bibnamefont {Moore}}, \bibinfo {author}
  {\bibfnamefont {G.}~\bibnamefont {Yusa}}, \bibinfo {author} {\bibfnamefont
  {T.}~\bibnamefont {Mano}}, \bibinfo {author} {\bibfnamefont {T.}~\bibnamefont
  {Noda}}, \bibinfo {author} {\bibfnamefont {M.}~\bibnamefont {Betz}},\ and\
  \bibinfo {author} {\bibfnamefont {A.~D.}\ \bibnamefont {Bristow}},\ }\href
  {https://doi.org/10.1103/PhysRevB.101.155414} {\bibfield  {journal} {\bibinfo
   {journal} {Phys. Rev. B}\ }\textbf {\bibinfo {volume} {101}},\ \bibinfo
  {pages} {155414} (\bibinfo {year} {2020})}\BibitemShut {NoStop}%
\bibitem [{\citenamefont {Kosaka}\ \emph {et~al.}(2008)\citenamefont {Kosaka},
  \citenamefont {Shigyou}, \citenamefont {Mitsumori}, \citenamefont {Rikitake},
  \citenamefont {Imamura}, \citenamefont {Kutsuwa}, \citenamefont {Arai},\ and\
  \citenamefont {Edamatsu}}]{Kosaka2008}%
  \BibitemOpen
  \bibfield  {author} {\bibinfo {author} {\bibfnamefont {H.}~\bibnamefont
  {Kosaka}}, \bibinfo {author} {\bibfnamefont {H.}~\bibnamefont {Shigyou}},
  \bibinfo {author} {\bibfnamefont {Y.}~\bibnamefont {Mitsumori}}, \bibinfo
  {author} {\bibfnamefont {Y.}~\bibnamefont {Rikitake}}, \bibinfo {author}
  {\bibfnamefont {H.}~\bibnamefont {Imamura}}, \bibinfo {author} {\bibfnamefont
  {T.}~\bibnamefont {Kutsuwa}}, \bibinfo {author} {\bibfnamefont
  {K.}~\bibnamefont {Arai}},\ and\ \bibinfo {author} {\bibfnamefont
  {K.}~\bibnamefont {Edamatsu}},\ }\href
  {https://doi.org/10.1103/PhysRevLett.100.096602} {\bibfield  {journal}
  {\bibinfo  {journal} {Phys. Rev. Lett.}\ }\textbf {\bibinfo {volume} {100}},\
  \bibinfo {pages} {096602} (\bibinfo {year} {2008})}\BibitemShut {NoStop}%
\bibitem [{\citenamefont {Kosaka}\ \emph {et~al.}(2009)\citenamefont {Kosaka},
  \citenamefont {Inagaki}, \citenamefont {Rikitake}, \citenamefont {Imamura},
  \citenamefont {Mitsumori},\ and\ \citenamefont {Edamatsu}}]{Kosaka2009}%
  \BibitemOpen
  \bibfield  {author} {\bibinfo {author} {\bibfnamefont {H.}~\bibnamefont
  {Kosaka}}, \bibinfo {author} {\bibfnamefont {T.}~\bibnamefont {Inagaki}},
  \bibinfo {author} {\bibfnamefont {Y.}~\bibnamefont {Rikitake}}, \bibinfo
  {author} {\bibfnamefont {H.}~\bibnamefont {Imamura}}, \bibinfo {author}
  {\bibfnamefont {Y.}~\bibnamefont {Mitsumori}},\ and\ \bibinfo {author}
  {\bibfnamefont {K.}~\bibnamefont {Edamatsu}},\ }\href
  {https://doi.org/10.1038/nature07729} {\bibfield  {journal} {\bibinfo
  {journal} {Nature}\ }\textbf {\bibinfo {volume} {457}},\ \bibinfo {pages}
  {702} (\bibinfo {year} {2009})}\BibitemShut {NoStop}%
\bibitem [{\citenamefont {Ren}\ \emph {et~al.}(2015)\citenamefont {Ren},
  \citenamefont {Kong}, \citenamefont {Li}, \citenamefont {Qian}, \citenamefont
  {Li}, \citenamefont {Tu},\ and\ \citenamefont {Wang}}]{Ren2015}%
  \BibitemOpen
  \bibfield  {author} {\bibinfo {author} {\bibfnamefont {Z.-C.}\ \bibnamefont
  {Ren}}, \bibinfo {author} {\bibfnamefont {L.-J.}\ \bibnamefont {Kong}},
  \bibinfo {author} {\bibfnamefont {S.-M.}\ \bibnamefont {Li}}, \bibinfo
  {author} {\bibfnamefont {S.-X.}\ \bibnamefont {Qian}}, \bibinfo {author}
  {\bibfnamefont {Y.}~\bibnamefont {Li}}, \bibinfo {author} {\bibfnamefont
  {C.}~\bibnamefont {Tu}},\ and\ \bibinfo {author} {\bibfnamefont {H.-T.}\
  \bibnamefont {Wang}},\ }\href {https://doi.org/10.1364/OE.23.026586}
  {\bibfield  {journal} {\bibinfo  {journal} {Opt. Express}\ }\textbf {\bibinfo
  {volume} {23}},\ \bibinfo {pages} {26586} (\bibinfo {year}
  {2015})}\BibitemShut {NoStop}%
\bibitem [{\citenamefont {Sato}\ \emph {et~al.}(shed)\citenamefont {Sato},
  \citenamefont {Matsumoto}, \citenamefont {Nakano}, \citenamefont {Ishihara},
  \citenamefont {Miyamoto}, \citenamefont {Omatsu},\ and\ \citenamefont
  {Morita}}]{Morita2022}%
  \BibitemOpen
  \bibfield  {author} {\bibinfo {author} {\bibfnamefont {S.}~\bibnamefont
  {Sato}}, \bibinfo {author} {\bibfnamefont {T.}~\bibnamefont {Matsumoto}},
  \bibinfo {author} {\bibfnamefont {Y.}~\bibnamefont {Nakano}}, \bibinfo
  {author} {\bibfnamefont {J.}~\bibnamefont {Ishihara}}, \bibinfo {author}
  {\bibfnamefont {K.}~\bibnamefont {Miyamoto}}, \bibinfo {author}
  {\bibfnamefont {T.}~\bibnamefont {Omatsu}},\ and\ \bibinfo {author}
  {\bibfnamefont {K.}~\bibnamefont {Morita}}} (\bibinfo {year}
  {unpublished})\BibitemShut {NoStop}%
\bibitem [{not()}]{note2022}%
  \BibitemOpen
  \bibinfo {note} {From the relation $\tilde{\beta}= -\gamma(\langle k_z^2
  \rangle - k_F^2/4)$, $\tilde{\beta}$ is estimated to be 1.44 meV$\text{\AA}$
  using the calculated value of $\langle k_z^2 \rangle =1.86\times 10^{16}$
  $\text{m}^{-2}$ and the reported bulk Dresselhaus coefficient of $\gamma
  =-11$ eV$\text{\AA}^3$. $|\alpha /\tilde{\beta}|$ is estimated from the
  experimental value of $-\alpha+\tilde{\beta}= 3.2$ meV$\text{\AA}$ and the
  estimated value of $\tilde{\beta}=1.44$ meV$\text{\AA}$.}\BibitemShut {Stop}%
\end{thebibliography}%

\end{document}